\documentclass[twocolumn,superscriptaddress,showpacs,preprintnumbers,amsmath,amssymb,aps,pre]{revtex4}
\usepackage{graphicx}
\usepackage{dcolumn}
\usepackage{bm}
\begin{document}
\title{Comments on the entropy of seismic electric signals under time reversal}
\author{P. A. Varotsos}
\email{pvaro@otenet.gr} \affiliation{Solid State Section and Solid
Earth Physics Institute, Physics Department, University of Athens,
Panepistimiopolis, Zografos 157 84, Athens, Greece}
\author{N. V. Sarlis}
\affiliation{Solid State Section and Solid Earth Physics
Institute, Physics Department, University of Athens,
Panepistimiopolis, Zografos 157 84, Athens, Greece}
\author{E. S. Skordas}
\affiliation{Solid State Section and Solid Earth Physics
Institute, Physics Department, University of Athens,
Panepistimiopolis, Zografos 157 84, Athens, Greece}
\author{M. S. Lazaridou}
\affiliation{Solid State Section and Solid Earth Physics
Institute, Physics Department, University of Athens,
Panepistimiopolis, Zografos 157 84, Athens, Greece}

\begin{abstract}
We present recent data of electric signals detected at the Earth's
surface, which confirm the earlier finding [\emph{Phys. Rev. E}
\textbf{73}, 031114 (2006)] that the value of the entropy in
natural time as well as its value under time reversal are smaller
than that of the entropy of a ``uniform'' distribution.
Furthermore, we show that the scale dependence of the fluctuations
of the natural time itself under time reversal provides a useful
tool for the discrimination of seismic electric signals (critical
dynamics) from  noises emitted from manmade sources as well as for
the determination of the scaling exponent.
\end{abstract}
\pacs{91.30.Dk, 05.40.-a, 05.45.Tp} \maketitle

In a time series comprising $N$ events, the natural time $\chi_{k}
= k/N$ serves as an index\cite{NAT01,NAT02} for the occurrence of
the $k$-th event. In natural time analysis, the time evolution of
the pair of the two quantities ($\chi_k, Q_k$) is considered,
where $Q_k$ denotes in general a quantity proportional to the
energy released during  the $k$-th event. In the case of
dichotomous electric signals (SES) activities (i.e., low frequency
$\leq 1$Hz electric signals that precede earthquakes, e.g., see
Refs.\cite{varbook,nat,newbook,VAR03b}) $Q_k$ stands for the
duration of the $k$-th pulse. The entropy $S$ in natural time is
defined\cite{NAT03B} as
\begin{equation}
S \equiv  \langle \chi \ln \chi \rangle - \langle \chi \rangle \ln
\langle \chi \rangle \label{Seq}
\end{equation}
where $\langle f( \chi) \rangle = \sum_{k=1}^N p_k f(\chi_k )$ and
$p_k=Q_{k}/\sum_{n=1}^{N}Q_{n}$. The value of the entropy upon
considering the time reversal ${\cal T}$, i.e., ${\cal T}
p_k=p_{N-k+1}$, is labelled by $S_-$.

SES activities (critical dynamics) exhibit infinitely ranged long
range temporal correlations\cite{NAT03,VAR06a,VAR06b} which are
destroyed\cite{VAR06b} after shuffling the durations $Q_k$
randomly. An interesting property emerged from the data analysis
of several SES activities refers to the fact\cite{VAR06a} that
both $S$ and $S_-$ values are smaller than the value of $S_u
(=1/2\ln 2-1/4\approx 0.0966$) of a ``uniform'' distribution
(defined in Refs. \cite{NAT01,NAT03B,NAT04,NAT05}, e.g.   when all
$p_k$ are equal), i.e.,
\begin{equation}
S, S_- < S_u
\end{equation}
This finding, which does {\em not} hold\cite{NAT05B} for
``artificial'' noises (AN) (i.e., electric signals emitted from
manmade sources), has been also supported by numerical
simulations, e.g., in fractional Brownian motion (fBm) time
series\cite{VAR06a} that have an exponent $\alpha_{DFA}$ resulted
from the Detrended Fluctuation Analysis (DFA)\cite{p18,p19} close
to unity. We clarify that fBm (with a self-similarity index
$H\approx1$) has been found\cite{WER05} as an appropriate type of
modelling process for the SES activities. Thus, it
seems\cite{VAR06a} that the validity of the relation (2) stems
from infinitely ranged long range correlations. The scope of this
paper is twofold: First, to provide the most recent experimental
data that strengthen the validity of the relation (2), and second
to point out the usefulness of the study of the fluctuations of
the natural time itself under time reversal.

Figure 1 depicts an electric signal, consisting of a number of
pulses, that has been recorded on November 14, 2006 at a station
labelled\cite{EPAPS} PIR lying in western Greece (close to Pirgos
city). This signal has been clearly collected at eleven measuring
electric dipoles with electrodes installed at sites that are
depicted in a map given in Ref.\cite{EPAPS}. The signal is
presented (continuous line in red) in Fig. 1(a) in normalized
units, i.e., by subtracting the mean value and dividing by the
standard deviation. (As for the actual amplitude -in $mV/km$- of
this SES activity\cite{EPAPS2}, it is comparable to the one
observed at the same station before the magnitude $M\approx7.0$
earthquake that occurred on Jan 8, 2006, see Ref.\cite{VAR06}.)
For the reader's convenience the corresponding dichotomous
representation is also drawn in Fig. 1(a) with a dotted (blue)
line, while in Fig. 1(c) we show (in red crosses) how the signal
is read in natural time. The computation of $S$ and $S_-$ leads to
the following values: $S=0.070\pm0.012$, $S_-=0.051\pm0.010$. As
for the variance\cite{NAT01,NAT02,NAT03,NAT03B} $\kappa_1\equiv
\langle \chi^2 \rangle - \langle \chi \rangle^2$, the resulting
value is $\kappa_1=0.062\pm0.010$. These values more or less obey
the conditions $\kappa_1\approx0.070$ and $S, S_- < S_u$ that have
been found to hold for other SES activities\cite{VAR06a}.

A closer inspection of Fig. 1(a) also reveals the following
experimental fact: An additional electric signal has been also
detected (in the gray shaded area of Fig. 1(a)), which consists of
pulses with markedly smaller amplitude than those of the SES
activity discussed in the previous paragraph. This is reproduced
(continuous line in red) in Fig. 1(b) in an expanded time scale
and for the sake of the reader's convenience its dichotomous
representation is also marked by the dotted (blue) line, which
leads to the natural time representation shown (dotted blue) in
Fig. 1(c). The computation of $S$ and $S_-$ gives
$S=0.077\pm0.004$, $S_-=0.082\pm0.004$, while $\kappa_1$ is found
to be $\kappa_1=0.076\pm0.005$. Hence, these values also obey the
aforementioned conditions ($\kappa_1\approx0.070$ and $S, S_- <
S_u$) for its classification as an SES activity.

\begin{figure}
\includegraphics{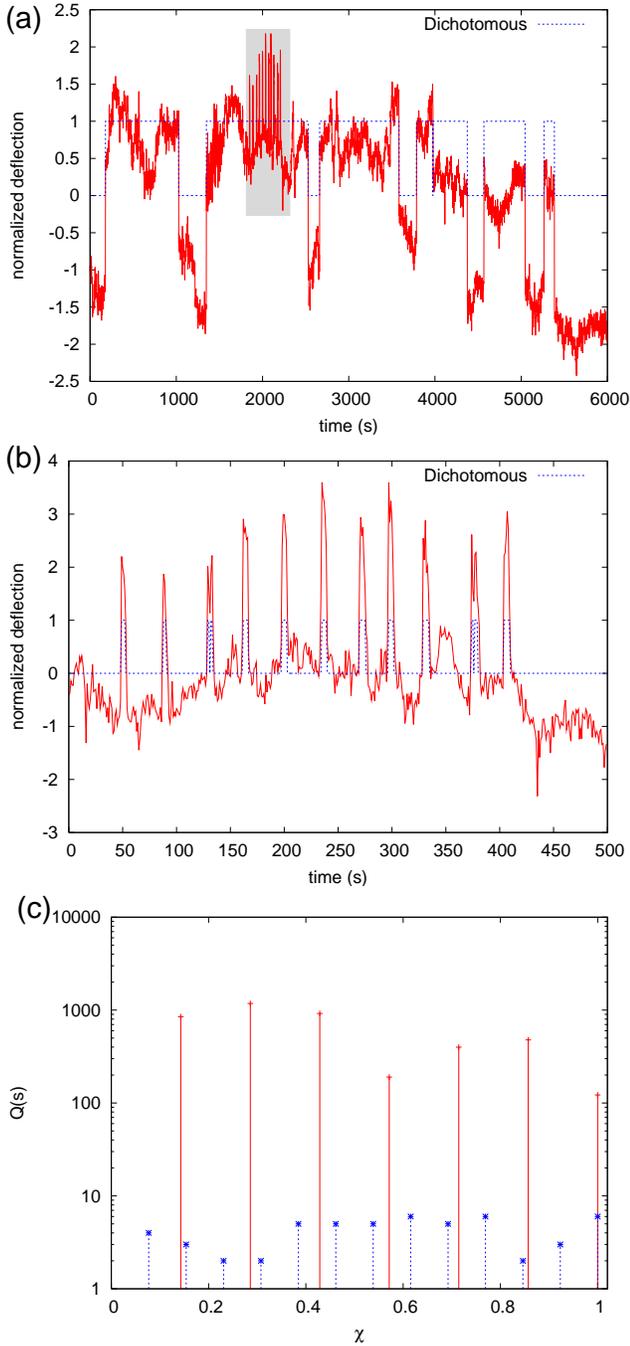}
\caption{\label{f1} (color online) (a) The electric signal
recorded on November 14, 2006 at PIR station (sampling rate
$f_{exp}=1$Hz). The signal is presented here in normalized units
(see the text). The corresponding dichotomous representation is
shown with the dotted (blue) line. The gray shaded area shows an
additional signal (consisting of pulses of smaller duration)
superimposed on the previously mentioned signal. (b) The signal
belonging to the gray shaded area in (a) is given here in an
expanded time scale, while its dichotomous representation is
marked by the dotted (blue) line. (c) How the signals in (a) and
(b) are read in natural time: the continuous (red crosses) and
dotted (blue asterisks) bars correspond to the durations of the
dichotomous representations marked in (a) and (b), respectively.}
\end{figure}

\begin{figure}
\includegraphics{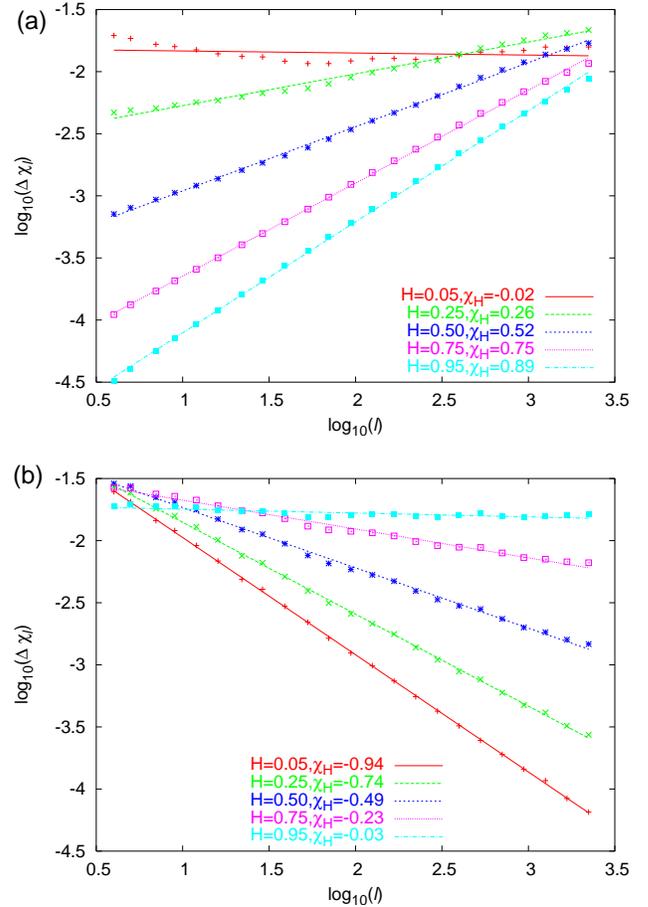}
\caption{\label{f2} (color online) The log-log plot of the
fluctuations $\Delta \chi_l$ of the natural time under time
reversal versus the scale $l$ for fBm (a) and fGn (b). Time series
of length $2\times10^4$ have been used.}
\end{figure}

\begin{figure}
\includegraphics{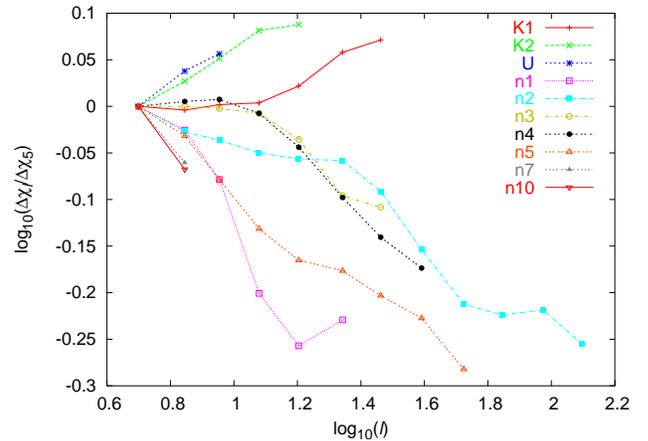}
\caption{\label{f3} (color online) The log-log plot of $\Delta
\chi_l$ versus the scale $l$ for three SES activities (K1, K2 and
U) and seven AN (n1-n5, n7 and n10) treated in Ref.\cite{NAT05B}
(cf., these signals have enough number of pulses in order to apply
the present analysis). The values of $\Delta \chi_l$ are divided
by the corresponding values $\Delta \chi _5$ at the scale $l=5$.}
\end{figure}

We now proceed to the second goal of this paper. The way through
which the entropy in natural time captures  the influence of the
effect of a small linear trend, has been studied in
Ref.\cite{VAR06a}. Namely, the ``uniform'' distribution,
$p(\chi)=1$, where $p(\chi )$ is a continuous probability density
function (PDF) corresponding to the point probabilities $p_k$ used
so far, has been perturbed by a linear trend $\epsilon$ in order
to obtain the parametric family of PDFs:
$p(\chi;\epsilon)=1+\epsilon (\chi -1/2)$. Such a family of PDFs
shares the interesting property ${\cal T}p(\chi;\epsilon)=
p(\chi;-\epsilon)$, i.e, the action of time reversal is obtained
by simply changing the sign of $\epsilon$. It has been
shown\cite{VAR06a} that the entropy $S(\epsilon )\equiv
S[p(\chi;\epsilon)]$, as well as that of the entropy under time
reversal $S_{-}(\epsilon )\equiv S[{\cal T}p(\chi;\epsilon )]$,
$S_{-}(\epsilon )=S(-\epsilon )$, depend \emph{non}-linearly on
the trend parameter $\epsilon$:
\begin{equation}
S(\epsilon)=-\frac{1}{4}+\frac{\epsilon}{72}-\left( \frac{1}{2}+\frac{\epsilon}{12} \right) \ln \left( \frac{1}{2}+\frac{\epsilon}{12} \right).
\end{equation}
However, it would be extremely useful to obtain a \emph{linear}
measure of $\epsilon$ in natural time. Actually, this is simply
the average of the natural time itself:
\begin{equation}
\langle \chi \rangle=\int_0^1 \chi p(\chi;\epsilon) d \chi =\frac{1}{2}+\frac{\epsilon}{12}.
\end{equation}
If we consider the fluctuations of this simple measure upon
time-reversal, we can obtain information on the long-range
dependence of $Q_k$. Actually, we shall show that a measure of the
long-range dependence emerges in natural time if we study the
dependence of its variance under time-reversal $\Delta \chi^2
\equiv  \text{E} [(\chi -{\cal T}\chi)^2 ]$ on the window length
$l$ that is used for the calculation. Since ${\cal
T}p_k=p_{l-k+1}$,  we have
\begin{equation}\label{e1}
  \Delta \chi^2 \equiv \text{E} [ (\chi -{\cal T}\chi)^2 ] = \text{E} \left\{ \left[ \sum_{k=1}^l \frac{k}{l} \left( p_k-p_{l-k+1} \right) \right]^2 \right\},
  \end{equation}
where the symbol $\text{E}[\ldots]$ denotes the expectation value
obtained when a window of length $l$ is sliding through the time
series $Q_k$.
  By expanding the square in the last part of Eq.(\ref{e1}), we obtain
\begin{widetext}
  \begin{equation}\label{e2}
   \Delta \chi^2 = \sum_{k=1}^l \left( \frac{k}{l}\right)^2 \text{E}[ (p_k-p_{l-k+1})^2 ] +
   \sum_{k\neq m} \frac{k m}{l^2} \text{E}[(p_k-p_{l-k+1}) (p_m-p_{l-m+1})].
  \end{equation}
The basic relation that interrelates $p_k$ is\cite{NAT04}
\begin{equation}\label{e0}
\sum_{k=1}^l p_k=1
\end{equation}
or equivalently $p_k=1-\sum_{m\neq k} p_m$. By subtracting from
the last expression, its value for  $k=l-k+1$, we obtain
$p_k-p_{l-k+1}=-\sum_{m\neq k} (p_m-p_{l-m+1})$, and thus
\begin{equation}\label{e3}
(p_k-p_{l-k+1})^2=-\sum_{m\neq k} (p_k-p_{l-k+1}) (p_m-p_{l-m+1}).
\end{equation}
By substituting Eq.(\ref{e3}) into Eq.(\ref{e2}), we obtain
\begin{equation}\label{e4}
\Delta \chi^2 = -\sum_{k=1}^l \left( \frac{k}{l}\right)^2
\sum_{m\neq k} \text{E} [ (p_k-p_{l-k+1}) (p_m-p_{l-m+1}) ] +
\sum_{k\neq m} \frac{k m}{l^2} \text{E}[ (p_k-p_{l-k+1})
(p_m-p_{l-m+1}) ]
\end{equation}
\end{widetext}
which simplifies to
\begin{equation}\label{e5}
\Delta \chi^2 = - \sum_{k,m} \frac{(k-m)^2}{l^2} \text{E}[
(p_k-p_{l-k+1}) (p_m-p_{l-m+1}) ]
\end{equation}
The negative sign appears because $(p_k-p_{l-k+1})$ and
$(p_m-p_{l-m+1})$ are in general anti-correlated due to
Eq.(\ref{e3}).  Equation (\ref{e5}) implies that $\Delta \chi^2$
measures the long-range correlations in $Q_k$: If we assume that
$-\text{E}[ (p_k-p_{l-k+1}) (p_m-p_{l-m+1})] \propto
(k-m)^{2\chi_H}/l^2$ (cf. $p_k$ scales as $1/l$, e.g. see
\cite{NAT04}), we have that $\Delta \chi^2 \propto
l^{4+2\chi_H}/l^4$ so that $\Delta \chi \propto l^{\chi_H}$ (where
$\chi_H$ is a scaling exponent). In order to examine the validity
of this result in the case when $Q_k$ are coming from a fractional
Gaussian noise (fGn) or fBm, we employed the following procedure:
First, we generated fBm (or fGn) time-series $X_k$ for a given
value of $H$ using the Mandelbrot-Weierstrass
function\cite{MAN69,SZU01,inter} as described in
Ref.\cite{VAR06a}. Second, since $Q_k$ should be positive, we
normalized the resulting $X_k$ time-series to zero mean and unit
standard deviation and then added to the normalized time-series
$N_k$ a constant factor $c$ to ensure the positivity of
$Q_k=N_k+c$ (for the purpose of the present study we used
$c=10.0$). The resulting $Q_k$ time-series were then analyzed and
the fluctuations of $\Delta \chi_l$ versus the scale $l$ are shown
in Figure 2. We observe that for fBm the estimator $\chi_H\approx
H$, whereas for fGn $\chi_H\approx H-1$. The physical meaning of
the present analysis was further investigated by performing the
same procedure in the time-series of the durations of the electric
signals analyzed in Ref.\cite{NAT05B}. The relevant results are
shown in Figure 3. Their inspection interestingly indicate that
all seven AN correspond to descending $\Delta \chi_l$ curves
versus the scale $l$, while the three SES to ascending curves.
This fact is in essential agreement with the results obtained in
Ref.\cite{NAT03}, which showed that the SES activities have a
stronger memory than AN. In other words, apart from being an
estimator of the scaling behavior, the ascending or descending
behavior of $\Delta \chi_l$ versus the scale $l$ seems to provide
a useful new tool for the classification of a signal as SES
activity or AN, respectively.


\begin{thebibliography}{22}
\expandafter\ifx\csname natexlab\endcsname\relax\def\natexlab#1{#1}\fi
\expandafter\ifx\csname bibnamefont\endcsname\relax
  \def\bibnamefont#1{#1}\fi
\expandafter\ifx\csname bibfnamefont\endcsname\relax
  \def\bibfnamefont#1{#1}\fi
\expandafter\ifx\csname citenamefont\endcsname\relax
  \def\citenamefont#1{#1}\fi
\expandafter\ifx\csname url\endcsname\relax
  \def\url#1{\texttt{#1}}\fi
\expandafter\ifx\csname urlprefix\endcsname\relax\def\urlprefix{URL }\fi
\providecommand{\bibinfo}[2]{#2}
\providecommand{\eprint}[2][]{\url{#2}}

\bibitem[{\citenamefont{Varotsos et~al.}(2001)\citenamefont{Varotsos, Sarlis,
  and Skordas}}]{NAT01}
\bibinfo{author}{\bibfnamefont{P.~A.} \bibnamefont{Varotsos}},
  \bibinfo{author}{\bibfnamefont{N.~V.} \bibnamefont{Sarlis}},
  \bibnamefont{and} \bibinfo{author}{\bibfnamefont{E.~S.}
  \bibnamefont{Skordas}}, \bibinfo{journal}{Practica of Athens Academy}
  \textbf{\bibinfo{volume}{76}}, \bibinfo{pages}{294} (\bibinfo{year}{2001}).

\bibitem[{\citenamefont{Varotsos et~al.}(2002)\citenamefont{Varotsos, Sarlis,
  and Skordas}}]{NAT02}
\bibinfo{author}{\bibfnamefont{P.~A.} \bibnamefont{Varotsos}},
  \bibinfo{author}{\bibfnamefont{N.~V.} \bibnamefont{Sarlis}},
  \bibnamefont{and} \bibinfo{author}{\bibfnamefont{E.~S.}
  \bibnamefont{Skordas}}, \bibinfo{journal}{Phys. Rev. E}
  \textbf{\bibinfo{volume}{66}}, \bibinfo{pages}{011902}
  (\bibinfo{year}{2002}).

\bibitem[{\citenamefont{Varotsos and Alexopoulos}(1986)}]{varbook}
\bibinfo{author}{\bibfnamefont{P.}~\bibnamefont{Varotsos}} \bibnamefont{and}
  \bibinfo{author}{\bibfnamefont{K.}~\bibnamefont{Alexopoulos}},
  \emph{\bibinfo{title}{Thermodynamics of Point Defects and their Relation with
  Bulk Properties}} (\bibinfo{publisher}{North Holland},
  \bibinfo{address}{Amsterdam}, \bibinfo{year}{1986}).

\bibitem[{\citenamefont{Varotsos et~al.}(1986)\citenamefont{Varotsos,
  Alexopoulos, Nomicos, and Lazaridou}}]{nat}
\bibinfo{author}{\bibfnamefont{P.}~\bibnamefont{Varotsos}},
  \bibinfo{author}{\bibfnamefont{K.}~\bibnamefont{Alexopoulos}},
  \bibinfo{author}{\bibfnamefont{K.}~\bibnamefont{Nomicos}}, \bibnamefont{and}
  \bibinfo{author}{\bibfnamefont{M.}~\bibnamefont{Lazaridou}},
  \bibinfo{journal}{Nature (London)} \textbf{\bibinfo{volume}{322}},
  \bibinfo{pages}{120} (\bibinfo{year}{1986}).

\bibitem[{\citenamefont{Varotsos}(2005)}]{newbook}
\bibinfo{author}{\bibfnamefont{P.}~\bibnamefont{Varotsos}},
  \emph{\bibinfo{title}{The Physics of Seismic Electric Signals}}
  (\bibinfo{publisher}{TERRAPUB}, \bibinfo{address}{Tokyo},
  \bibinfo{year}{2005}).

\bibitem[{\citenamefont{Varotsos
  et~al.}(2003{\natexlab{a}})\citenamefont{Varotsos, Sarlis, and
  Skordas}}]{VAR03b}
\bibinfo{author}{\bibfnamefont{P.}~\bibnamefont{Varotsos}},
  \bibinfo{author}{\bibfnamefont{N.}~\bibnamefont{Sarlis}}, \bibnamefont{and}
  \bibinfo{author}{\bibfnamefont{S.}~\bibnamefont{Skordas}},
  \bibinfo{journal}{Phys. Rev. Lett.} \textbf{\bibinfo{volume}{91}},
  \bibinfo{pages}{148501} (\bibinfo{year}{2003}{\natexlab{a}}).

\bibitem[{\citenamefont{Varotsos
  et~al.}(2003{\natexlab{b}})\citenamefont{Varotsos, Sarlis, and
  Skordas}}]{NAT03B}
\bibinfo{author}{\bibfnamefont{P.~A.} \bibnamefont{Varotsos}},
  \bibinfo{author}{\bibfnamefont{N.~V.} \bibnamefont{Sarlis}},
  \bibnamefont{and} \bibinfo{author}{\bibfnamefont{E.~S.}
  \bibnamefont{Skordas}}, \bibinfo{journal}{Phys. Rev. E}
  \textbf{\bibinfo{volume}{68}}, \bibinfo{pages}{031106}
  (\bibinfo{year}{2003}{\natexlab{b}}).

\bibitem[{\citenamefont{Varotsos
  et~al.}(2003{\natexlab{c}})\citenamefont{Varotsos, Sarlis, and
  Skordas}}]{NAT03}
\bibinfo{author}{\bibfnamefont{P.~A.} \bibnamefont{Varotsos}},
  \bibinfo{author}{\bibfnamefont{N.~V.} \bibnamefont{Sarlis}},
  \bibnamefont{and} \bibinfo{author}{\bibfnamefont{E.~S.}
  \bibnamefont{Skordas}}, \bibinfo{journal}{Phys. Rev. E}
  \textbf{\bibinfo{volume}{67}}, \bibinfo{pages}{021109}
  (\bibinfo{year}{2003}{\natexlab{c}}).

\bibitem[{\citenamefont{Varotsos
  et~al.}(2006{\natexlab{a}})\citenamefont{Varotsos, Sarlis, Skordas, Tanaka,
  and Lazaridou}}]{VAR06a}
\bibinfo{author}{\bibfnamefont{P.~A.} \bibnamefont{Varotsos}},
  \bibinfo{author}{\bibfnamefont{N.~V.} \bibnamefont{Sarlis}},
  \bibinfo{author}{\bibfnamefont{E.~S.} \bibnamefont{Skordas}},
  \bibinfo{author}{\bibfnamefont{H.~K.} \bibnamefont{Tanaka}},
  \bibnamefont{and} \bibinfo{author}{\bibfnamefont{M.~S.}
  \bibnamefont{Lazaridou}}, \bibinfo{journal}{Phys. Rev. E}
  \textbf{\bibinfo{volume}{73}}, \bibinfo{pages}{031114}
  (\bibinfo{year}{2006}{\natexlab{a}}).

\bibitem[{\citenamefont{Varotsos
  et~al.}(2006{\natexlab{b}})\citenamefont{Varotsos, Sarlis, Skordas, Tanaka,
  and Lazaridou}}]{VAR06b}
\bibinfo{author}{\bibfnamefont{P.~A.} \bibnamefont{Varotsos}},
  \bibinfo{author}{\bibfnamefont{N.~V.} \bibnamefont{Sarlis}},
  \bibinfo{author}{\bibfnamefont{E.~S.} \bibnamefont{Skordas}},
  \bibinfo{author}{\bibfnamefont{H.~K.} \bibnamefont{Tanaka}},
  \bibnamefont{and} \bibinfo{author}{\bibfnamefont{M.~S.}
  \bibnamefont{Lazaridou}}, \bibinfo{journal}{Phys. Rev. E}
  \textbf{\bibinfo{volume}{74}}, \bibinfo{pages}{021123}
  (\bibinfo{year}{2006}{\natexlab{b}}).

\bibitem[{\citenamefont{Varotsos et~al.}(2004)\citenamefont{Varotsos, Sarlis,
  Skordas, and Lazaridou}}]{NAT04}
\bibinfo{author}{\bibfnamefont{P.~A.} \bibnamefont{Varotsos}},
  \bibinfo{author}{\bibfnamefont{N.~V.} \bibnamefont{Sarlis}},
  \bibinfo{author}{\bibfnamefont{E.~S.} \bibnamefont{Skordas}},
  \bibnamefont{and} \bibinfo{author}{\bibfnamefont{M.~S.}
  \bibnamefont{Lazaridou}}, \bibinfo{journal}{Phys. Rev. E}
  \textbf{\bibinfo{volume}{70}}, \bibinfo{pages}{011106}
  (\bibinfo{year}{2004}).

\bibitem[{\citenamefont{Varotsos
  et~al.}(2005{\natexlab{a}})\citenamefont{Varotsos, Sarlis, Skordas, and
  Lazaridou}}]{NAT05}
\bibinfo{author}{\bibfnamefont{P.~A.} \bibnamefont{Varotsos}},
  \bibinfo{author}{\bibfnamefont{N.~V.} \bibnamefont{Sarlis}},
  \bibinfo{author}{\bibfnamefont{E.~S.} \bibnamefont{Skordas}},
  \bibnamefont{and} \bibinfo{author}{\bibfnamefont{M.~S.}
  \bibnamefont{Lazaridou}}, \bibinfo{journal}{Phys. Rev. E}
  \textbf{\bibinfo{volume}{71}}, \bibinfo{pages}{011110}
  (\bibinfo{year}{2005}{\natexlab{a}}).

\bibitem[{\citenamefont{Varotsos
  et~al.}(2005{\natexlab{b}})\citenamefont{Varotsos, Sarlis, Tanaka, and
  Skordas}}]{NAT05B}
\bibinfo{author}{\bibfnamefont{P.~A.} \bibnamefont{Varotsos}},
  \bibinfo{author}{\bibfnamefont{N.~V.} \bibnamefont{Sarlis}},
  \bibinfo{author}{\bibfnamefont{H.~K.} \bibnamefont{Tanaka}},
  \bibnamefont{and} \bibinfo{author}{\bibfnamefont{E.~S.}
  \bibnamefont{Skordas}}, \bibinfo{journal}{Phys. Rev. E}
  \textbf{\bibinfo{volume}{71}}, \bibinfo{pages}{032102}
  (\bibinfo{year}{2005}{\natexlab{b}}).

\bibitem[{\citenamefont{Peng et~al.}(1994)\citenamefont{Peng, Buldyrev, Havlin,
  Simons, Stanley, and Goldberger}}]{p18}
\bibinfo{author}{\bibfnamefont{C.-K.} \bibnamefont{Peng}},
  \bibinfo{author}{\bibfnamefont{S.~V.} \bibnamefont{Buldyrev}},
  \bibinfo{author}{\bibfnamefont{S.}~\bibnamefont{Havlin}},
  \bibinfo{author}{\bibfnamefont{M.}~\bibnamefont{Simons}},
  \bibinfo{author}{\bibfnamefont{H.~E.} \bibnamefont{Stanley}},
  \bibnamefont{and} \bibinfo{author}{\bibfnamefont{A.~L.}
  \bibnamefont{Goldberger}}, \bibinfo{journal}{Phys. Rev. E}
  \textbf{\bibinfo{volume}{49}}, \bibinfo{pages}{1685} (\bibinfo{year}{1994}).

\bibitem[{\citenamefont{Buldyrev et~al.}(1995)\citenamefont{Buldyrev,
  Goldberger, Havlin, Mantegna, Matsa, Peng, Simons, and Stanley}}]{p19}
\bibinfo{author}{\bibfnamefont{S.~V.} \bibnamefont{Buldyrev}},
  \bibinfo{author}{\bibfnamefont{A.~L.} \bibnamefont{Goldberger}},
  \bibinfo{author}{\bibfnamefont{S.}~\bibnamefont{Havlin}},
  \bibinfo{author}{\bibfnamefont{R.~N.} \bibnamefont{Mantegna}},
  \bibinfo{author}{\bibfnamefont{M.~E.} \bibnamefont{Matsa}},
  \bibinfo{author}{\bibfnamefont{C.-K.} \bibnamefont{Peng}},
  \bibinfo{author}{\bibfnamefont{M.}~\bibnamefont{Simons}}, \bibnamefont{and}
  \bibinfo{author}{\bibfnamefont{H.~E.} \bibnamefont{Stanley}},
  \bibinfo{journal}{Phys. Rev. E} \textbf{\bibinfo{volume}{51}},
  \bibinfo{pages}{5084} (\bibinfo{year}{1995}).

\bibitem[{\citenamefont{Weron et~al.}(2005)\citenamefont{Weron, Burnecki,
  Mercik, and Weron}}]{WER05}
\bibinfo{author}{\bibfnamefont{A.}~\bibnamefont{Weron}},
  \bibinfo{author}{\bibfnamefont{K.}~\bibnamefont{Burnecki}},
  \bibinfo{author}{\bibfnamefont{S.}~\bibnamefont{Mercik}}, \bibnamefont{and}
  \bibinfo{author}{\bibfnamefont{K.}~\bibnamefont{Weron}},
  \bibinfo{journal}{Phys. Rev. E} \textbf{\bibinfo{volume}{71}},
  \bibinfo{pages}{016113} (\bibinfo{year}{2005}).

\bibitem[{EPA({\natexlab{a}})}]{EPAPS}
\eprint{See EPAPS Document No. E-PLEEE8-74-190608 for additional information,
  originally from P.A. Varotsos, N.V. Sarlis, E.S. Skordas, H.K. Tanaka and
  M.S. Lazaridou, Phys. Rev. E 74, 021123 (2006). For more information on
  EPAPS, see http://www.aip.org/pubservs/ epaps.html.}

\bibitem[{EPA({\natexlab{b}})}]{EPAPS2}
\eprint{The time of the impending earthquake can be determined by means of the
  procedure described in EPAPS Document No. E-PLEEE8-73-134603 for additional
  information, originally from P.A. Varotsos, N.V. Sarlis, E.S. Skordas, H.K.
  Tanaka and M.S. Lazaridou, Phys. Rev. E 73, 031114 (2006). For more
  information on EPAPS, see http://www.aip.org/pubservs/ epaps.html.}

\bibitem[{\citenamefont{Varotsos}(2006)}]{VAR06}
\bibinfo{author}{\bibfnamefont{P.~A.} \bibnamefont{Varotsos}},
  \bibinfo{journal}{Proc. Japan Acad., Ser. B} \textbf{\bibinfo{volume}{82}},
  \bibinfo{pages}{86} (\bibinfo{year}{2006}).

\bibitem[{\citenamefont{Mandelbrot and Wallis}(1969)}]{MAN69}
\bibinfo{author}{\bibfnamefont{B.}~\bibnamefont{Mandelbrot}} \bibnamefont{and}
  \bibinfo{author}{\bibfnamefont{J.~R.} \bibnamefont{Wallis}},
  \bibinfo{journal}{Water Resources Research} \textbf{\bibinfo{volume}{5}},
  \bibinfo{pages}{243} (\bibinfo{year}{1969}).

\bibitem[{\citenamefont{Szulga and Molz}(2001)}]{SZU01}
\bibinfo{author}{\bibfnamefont{J.}~\bibnamefont{Szulga}} \bibnamefont{and}
  \bibinfo{author}{\bibfnamefont{F.}~\bibnamefont{Molz}}, \bibinfo{journal}{J.
  Stat. Phys.} \textbf{\bibinfo{volume}{104}}, \bibinfo{pages}{1317}
  (\bibinfo{year}{2001}).

\bibitem[{\citenamefont{Frame et~al.}()\citenamefont{Frame, Mandelbrot, and
  Neger}}]{inter}
\bibinfo{author}{\bibfnamefont{M.}~\bibnamefont{Frame}},
  \bibinfo{author}{\bibfnamefont{B.}~\bibnamefont{Mandelbrot}},
  \bibnamefont{and} \bibinfo{author}{\bibfnamefont{N.}~\bibnamefont{Neger}},
  \bibinfo{note}{fractal Geometry, Yale University, available from
  \url{http://classes.yale.edu/fractals/}, see
  \url{http://classes.yale.edu/Fractals/RandFrac/fBm/fBm4.html}}.

\end{thebibliography}
\bibliographystyle{apsrev}

\end{document}